\documentclass[twocolumn]{aastex61}
\usepackage{amsmath}

\submitjournal{ApJ}

\shorttitle{TTV modes}
\shortauthors{Linial, Gilbaum \& Sari}

\begin{document}

\title{Modal Decomposition of TTV - Inferring Planet Masses and Eccentricities}


\correspondingauthor{Itai Linial}
\email{itai.linial@mail.huji.ac.il}

\author{Itai Linial}
\affil{Racah Institute of Physics, The Hebrew University, Jerusalem 91904, Israel}

\author{Shmuel Gilbaum}
\affil{Racah Institute of Physics, The Hebrew University, Jerusalem 91904, Israel}

\author{Re'em Sari}
\affiliation{Racah Institute of Physics, The Hebrew University, Jerusalem 91904, Israel}

\begin{abstract}

Transit timing variations (TTVs) are a powerful tool for characterizing the properties of transiting exoplanets. However, inferring planet properties from the observed timing variations is a challenging task, which is usually addressed by extensive numerical searches. We propose a new, computationally inexpensive method for inverting TTV signals in a planetary system of two transiting planets. To the lowest order in planetary masses and eccentricities, TTVs can be expressed as a linear combination of 3 functions, which we call the \textit{TTV modes}. These functions depend only on the planets' linear ephemerides, and can be either constructed analytically, or by performing 3 orbital integrations of the three-body system. Given a TTV signal, the underlying physical parameters are found by decomposing the data as a sum of the TTV modes. We demonstrate the use of this method by inferring the mass and eccentricity of 6 \textit{Kepler} planets that were previously characterized in other studies. Finally we discuss the implications and future prospects of our new method.

\end{abstract}



\section{Introduction} \label{sec:intro}

When a transiting exoplanet is orbiting in the presence of a non-Keplerian potential, the timing of its transits may deviate from constant periodicity. Such deviations from linear ephemeris, known as transit timing variations (TTVs), possess valuable information regarding the perturbations to the Keplerian potential.

A primary source of TTVs is the presence of additional massive planets orbiting the same host star. TTVs were therefore recognized as a method for detecting non-transiting planets in planetary systems with at least one transiting planet and as a tool for confirming planet candidates \citep{Miralda_Escude_2002,Agol_2005,Holman_2005,Nesvorny_2012, Fabrycky_2012}.

TTVs can also serve as a proxy for constraining the masses and orbital parameters of the perturbing planets. When radial velocity measurements of the host star are not available, TTVs are the primary tool for constraining the mass (and bulk density) of transiting planets.

The \textit{Kepler} mission has provided TTV measurements for hundreds of planets \citep{Rowe_Catalog_2015,Holczer_Catalog_2016}. Several authors have calculated planet masses and eccentricities of \textit{Kepler} planets using the released data \citep{HL_2014, HL_2016,HL_2017, Lissauer_2013}. Most commonly, the inversion of TTVs for inferring planet properties is done using Markov chain Monte Carlo (MCMC) simulations \citep{HL_2016,HL_2017,JH_2016}. MCMC simulations sample planet masses and orbital parameters, searching for the model most consistent with the observations. Typically these procedures require computation of $10^7$ - $10^9$ N-body orbital integrations, each lasting for a few hundreds of orbits \citep{Deck_TTVFast_2014}. These intensive numerical searches are often complemented by fitting to approximate TTV formulae, that help in identifying the parameter degeneracies within the numerical searches \citep{HL_2016,HL_2017,JH_2016,AD_first_2016}.

In this work, we propose a computationally inexpensive method for inferring planet properties from TTV signals of systems with two transiting planets. Assuming TTV amplitudes which are small compared to the orbital periods, low planet-to-star mass ratios and small orbital eccentricities, TTVs can be expressed as a linear sum of 3 dimensionless functions which we call the \textit{TTV modes}. The TTV modes depend solely on the planets' period ratio and orbital phase, which are known \textit{a-priori}. We demonstrate how these modes can be easily constructed with merely 3 orbital integrations. Given a TTV signal, it can be decomposed to a sum of these modes, with coefficients that determine the unknown physical parameters of the perturbing planet.

Using data from TTV catalogs \citep{Rowe_Catalog_2015, Holczer_Catalog_2016} we calculate the mass and eccentricity of 6 \textit{Kepler} planets, (Kepler-307 b \& c, Kepler-177 b \& c, Kepler-11 d \& e). The masses we infer are consistent with the values reported by previous works. 

The outline of the paper is as follows. In section \ref{sec:TTV_Theory} we present the general form that TTVs take, and define the TTV modes. We discuss the procedure of fitting the modes and inferring the physical parameters in section \ref{sec:Mode_recon}. In section \ref{sec:Results} we demonstrate the use of our method on a set of 6 \textit{Kepler} planets. We discuss the interpretation of the TTV modes and compare them to other works in section \ref{sec:Interpreting_the_ttv_modes}. Finally, we discuss our conclusions and future prospects in section \ref{sec:Discussion}.

\section{TTV Theory} \label{sec:TTV_Theory}

\subsection{Leading order TTV} \label{sec:Leading_order_TTV}
Consider a planetary system with two transiting planets. Small mutual inclinations tend to have a negligible effect on TTVs \citep{NV_2014,HL_2016,AD_first_2016}. Additionally, systems with multiple transiting planets have been observationally shown to be nearly coplanar, with typical mutual inclination of less than a few degrees \citep{Figueira_2012,Tremaine_Dong_2012,Fabrycky_2014}. We will therefore assume that the two planets are on coplanar orbits. The TTVs of these planets depend on their masses and eccentricities. We define two small parameters, $\mu = M_p/M_\star$ and $\mu' = M_p'/M_\star$, the relative masses of the inner and outer planet. 

TTVs are insensitive to the individual eccentricities of the interacting planets, but rather to the \textit{combined free eccentricity} vector \citep{LXW_2012,HL_2016,HL_2017} - a weighted difference between the free-eccentricities of the two planets (see appendix \ref{sec:Appendix_free_eccentricities}). The dependence of TTVs on this combined quantity rather than the individual eccentricities is known as the eccentricity-eccentricity degeneracy (see \cite{LXW_2012, JH_2016} for further discussion). 

One exception where this degeneracy may be lifted, is the case of proximity to the $2:1$ mean-motion resonance (MMR). A few studies have reported eccentricity measurements of individual planets in systems close to the $2:1$ MMR \citep{Holman_2010, Borsato_2014}. Excluding this case, TTVs strongly depend only on the system's combined free eccentricity.

We define $e_x$, the combined eccentricity component parallel to the observer's line of sight, and $e_y$, the component perpendicular to the observer (in the positive sense of rotation).

To the leading order in these small parameters, the TTV of the inner planet can be written in the following form
\begin{equation} \label{eq:gen_ttv_linear}
	ttv_i = \mu' P \left[ G_i + e_x F_i^x + e_y F_i^y \right] \,,
\end{equation}
where $G$, $F^x$ and $F^y$ are functions of the period ratio, $P'/P$ and the angle between the planets at the time of the first transit, $\Delta \theta_0$ - both values are assumed to be known. We refer to these functions as the \textit{TTV modes}. The TTV of the outer planet takes a similar functional form,
\begin{equation} \label{eq:gen_ttv_linear_outer}
	ttv'_i = \mu P' \left[ g_i + e_x f_i^{x} + e_y f_i^{y} \right] \,,
\end{equation}
where $g$, $f^{x}$ and $f^{y}$ are also functions of $P'/P$ and $\Delta \theta_0$. Note that the TTV of the inner planet depends on the mass of its perturbing companion, $\mu'$, and vice versa. 

The general form TTVs take is given in equations \ref{eq:gen_ttv_linear} and \ref{eq:gen_ttv_linear_outer}. TTVs vanish when planetary masses go to zero, as the Keplerian orbits remain unperturbed. Since TTVs occur due to the gravitational interaction with a perturber, TTVs are expected to change linearly with mass, to lowest order, and hence the expressions are proportional to $\mu'$ and $\mu$.

TTVs depend mostly on the combined free eccentricity vector, rather than the individual eccentricities of the two planets. This behavior was shown both analytically and numerically by several authors \citep{HL_2016,JH_2016}. Intuitively, this can be understood by the following argument. Assuming small individual eccentricities, an elliptical orbit, to first order in eccentricity, is given by a circle, with its center shifted by an amount proportional to its eccentricity, in the direction of its apocenter. Since the gravitational interaction between the planets depends on their relative positions, eccentricity enters through roughly as the difference in their eccentricity vectors. For instance, if two nearby planets have the same eccentricity vector, their relative motion will be along two concentric circular orbits, eliminating the effect of eccentricity.

The functions $G$, $F^x$ and $F^y$ (and respectively, $g$, $f^x$ and $f^y$) are linearly independent, and span the TTV space. Given the inner planet's TTV signal $ttv_i$, the unknown parameters $\mu'$, $e_x$ and $e_y$ can be uniquely determined by projecting $ttv_i$ onto the basis functions $G$, $F^x$ and $F^y$. 

If the TTV signal contains no noise, the error in the parameter inference is strictly due to higher order terms not included in equation \ref{eq:gen_ttv_linear}, $e_x^2$, $e_y^2$, $\mu'\mu$ etc. Since the planet to star mass ratios and orbital eccentricities are quite small, these will generally account for small corrections. An exception is when the planets are near a second-order MMR, where second-order terms in eccentricity dominate (see the TTV expressions in \cite{HL_2016,HL_2017} that account for higher order terms).

While the functions $G$, $F^x$ and $F^y$ are linearly independent, they do not form an orthogonal basis. Therefore, using equation \ref{eq:gen_ttv_linear} to fit a noisy signal would result in correlated errors in the coefficients of the functions $G$, $F^x$ and $F^y$, and accordingly in the inferred physical parameters. 

We note that when a system is near a first-order mean-motion resonance, $G$ and $F^x$ are nearly parallel (see for example figure \ref{fig:Modes_x_0_3}). This results in the so-called mass-eccentricity degeneracy \citep{LXW_2012,HL_2014,HL_2016}. When the data is too noisy to disentangle the contributions of $G$ and $F^x$, the strong correlation between $\mu'$ and $\mu' e_x$ poses a difficulty in the parameter inference.

The TTV modes are easily obtained - either using analytical formulae or numerically, constructing the functions from a small number of N-body integrations (see section \ref{sec:contruct_modes}). The parameter inference problem reduces to finding the three parameters $\mu'$, $e_x$ and $e_y$ that minimize the error given the data samples. Since the theoretical TTV expression (right hand side of equation \ref{eq:gen_ttv_linear}) is linear in $\mu'$, $\mu' e_x$ and $\mu' e_y$, the best-fit parameters are obtained analytically through a single three-by-three matrix inversion (see section \ref{sec:Mode_recon}). 

\subsection{Constructing the TTV modes} \label{sec:contruct_modes}
The TTV modes, $G$, $F^x$ and $F^y$ can be easily constructed by running 3 numerical orbital integrations. The modes are functions of $P'/P$ and $\Delta \theta_0$ - the relative angular positions at the the time of first transit. Three N-body integrations are performed, all consistent with the target $P'/P$ and $\Delta \theta_0$. We set a small value for the two masses  $\mu_0$ and $\mu'_0$. These masses have to be sufficiently small such that the TTVs are in the linear regime, yet large enough that the TTV signal is easily measured from the simulation. In practice we have used  $\mu_0=\mu'_0=10^{-7}$ ($\ll M_{\oplus} / M_{\odot}$). With these masses and phases, we run the following set of simulations, and measure the TTV for the inner planet:

\begin{enumerate}
\item Both planets possess zero free-eccentricities. We denote the TTV of this simulation by $ttv_i^0$.
\item The inner planet has a small free-eccentricity $e_{x0}$, with the pericenter facing the observer, while the outer planet has zero free-eccentricity. We have used $e_{x0} = 10^{-3}$. We denote the TTV calculated from this simulation as $ttv_i^x$.
\item The inner planet has a small free-eccentricity $e_{y0}$, with the pericenter perpendicular to the observer, while the outer planet has zero free-eccentricity. We have used $e_{y0} = 10^{-3}$. We denote the TTV calculated from this simulation as $ttv_i^y$.
\end{enumerate}

Some of the intermediate transit indices may be missing from the observational data. Thus, when calculating the TTVs from the simulation results, we limit ourselves to the transit indices that appear in the data. Therefore, the modes also depend on the list of transit indices available in the data, rather than just $P'/P$ and $\Delta \theta_0$.

Note that setting the correct initial conditions that yield a specific free-eccentricity requires some care. In appendices \ref{sec:Appendix_free_eccentricities} and \ref{sec:Appendix_Initial_conditions_free_eccentricity} we provide an analytical expression for the resultant free-eccentricity as a function of the initial eccentricity. 

Given these three synthetic TTV signals, we obtain the TTV modes:

\begin{equation}
	G_i =  ttv_i^0 \frac{1}{\mu'_0 P_0} \,,
\end{equation}

\begin{equation}
	F_i^x = \left( ttv_i^x - ttv_i^0 \right) \frac{1}{\mu'_0 P_0 e_{x0}} \,,
\end{equation}

\begin{equation}
	F_i^y = \left( ttv_i^y - ttv_i^0 \right) \frac{1}{\mu'_0 P_0 e_{y0}} \,,
\end{equation}
where $P_0$ is the period of the inner planet in the numerical integration. The modes may be calculated with any set of three independent simulations, i.e., with different pericenter orientations. More generally, the TTV modes could also be calculated by running  $N \geq 3$ simulations. Then, for each $i$ separately, find the values of $F^x_i$, $F^y_i$ \& $G_i$ that best solve the set of $N$ equations (\ref{eq:gen_ttv_linear}), in the $\chi^2$ sense. 

\begin{figure}[!ht]
\includegraphics[width=1.04\columnwidth]{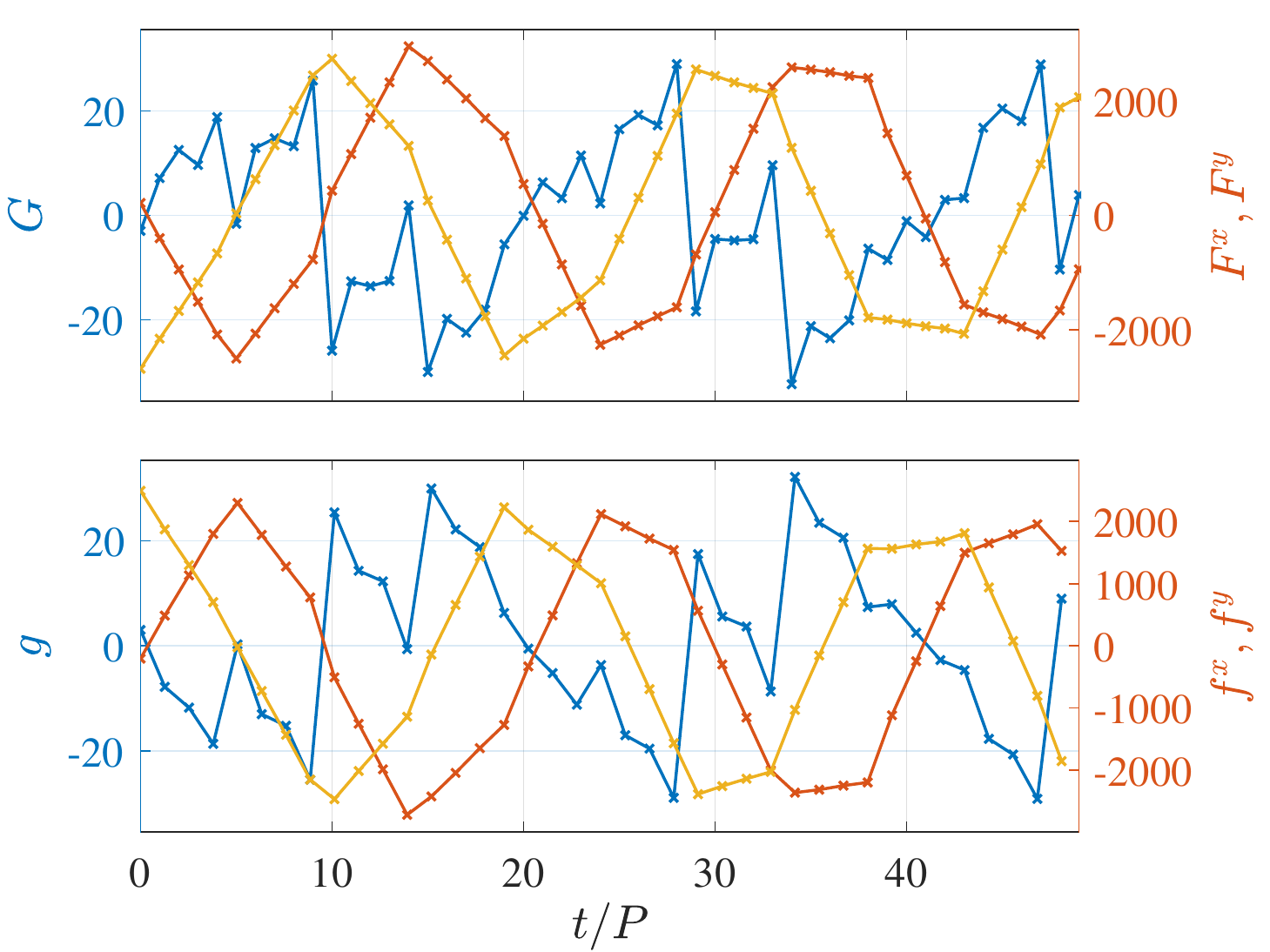}
\caption{TTV modes, calculated for $P'/P = 1.265$, and $\Delta \theta_0 = 9\pi/10$. \textit{Top panel} - the inner planet's TTV modes, \textit{bottom panel} - the outer planet's TTV modes. Left hand vertical axis - the value of $G$, right hand - the value of $F^x$ and $F^y$. The \textit{blue curve} is $G$, the zero eccentricity mode. The \textit{red curve} is $F^x$, the $e_x$ mode. The \textit{yellow curve} is $F^y$, the $e_y$ mode. 
\label{fig:Modes_x_0_17}}
\end{figure}

\begin{figure}[!ht]
\includegraphics[width=1.04\columnwidth]{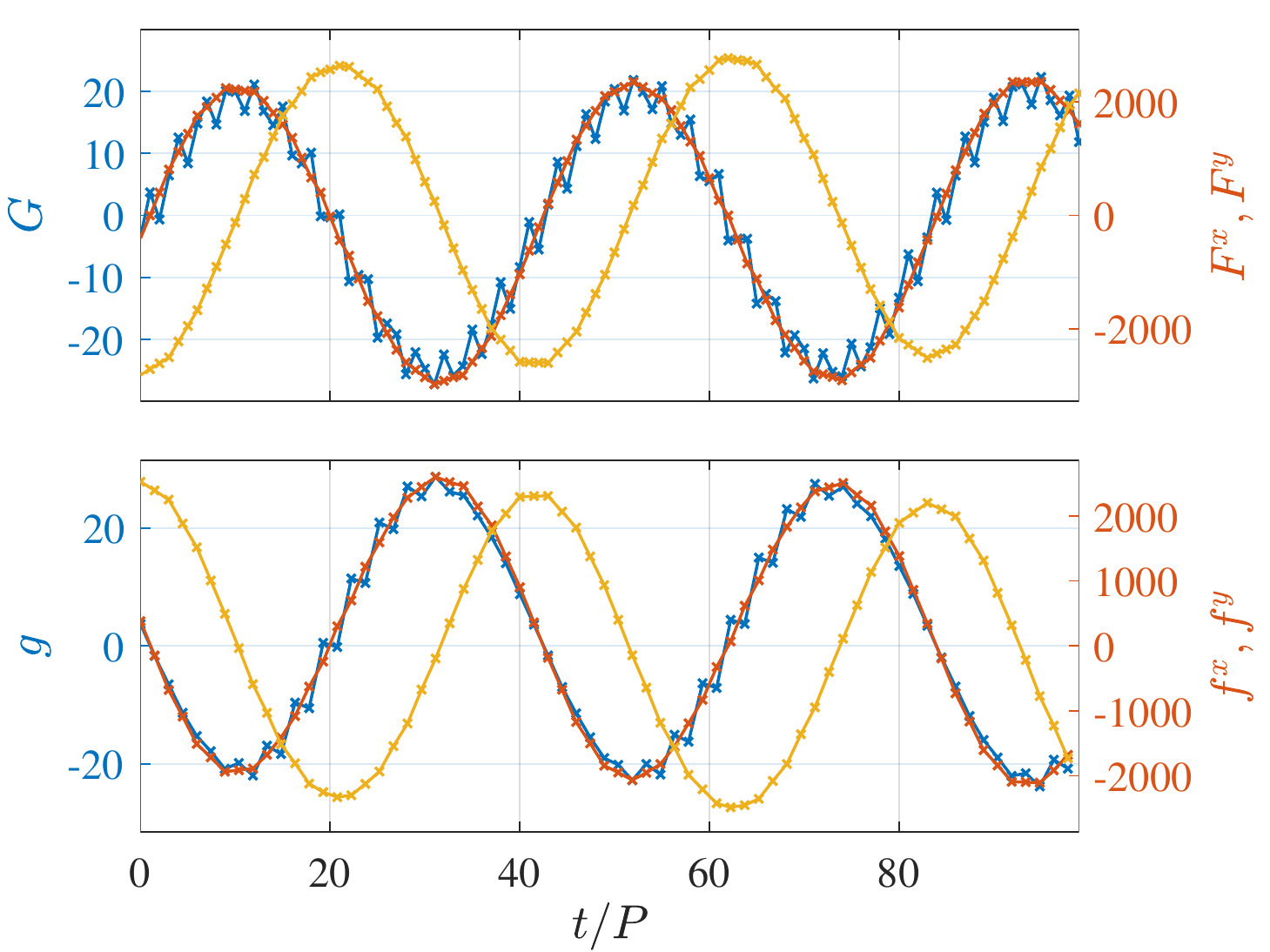}
\caption{TTV modes, calculated for $P'/P = 1.482$, and $\Delta \theta_0 = 9\pi/10$. \textit{Top panel} - the inner planet's TTV modes, \textit{bottom panel} - the outer planet's TTV modes. Left hand vertical axis - the value of $G$, right hand - the value of $F^x$ and $F^y$. The \textit{red curve} is $F^x$, the $e_x$ mode. The \textit{yellow curve} is $F^y$, the $e_y$ mode. $F^x$ and $F^y$ are roughly sinusoidal, and have a phase difference of $\pi/2$. Note that $G$ and $F^x$ are nearly parallel.\label{fig:Modes_x_0_3}}
\end{figure}

Figures \ref{fig:Modes_x_0_17} and \ref{fig:Modes_x_0_3} demonstrate the TTV modes associated with different period ratios, calculated using the above procedure. In figure \ref{fig:Modes_x_0_17}, $P'/P=1.265$ and $\Delta \theta_0 = 9 \pi / 10$. This orbital configuration lies between the the $5:4$ and $4:3$ MMRs. 

Figure \ref{fig:Modes_x_0_3} demonstrates the TTV modes corresponding to a period ratio $P'/P = 1.482$, and an initial angular separation $\Delta \theta_0 = 9 \pi/10$. Due to system's proximity to the $3:2$ MMR, the modes can be interpreted in terms of the TTV expressions of \cite{HL_2016, HL_2017} (see section \ref{sec:Interpreting_the_ttv_modes} for further discussion on this comparison). $F^x$ and $F^y$ modes are roughly sinusoidal, slowly varying over the superperiod, similar to the 'fundamental' TTV term (as defined in equation 1 of \cite{HL_2017}). The $G$ mode contains the 'chopping' TTV term (see \cite{DA_Conj_2015,HL_2016}), with a distinct synodic frequency. This figure also demonstrates the near-parallel nature of $G$ and $F^x$. Up to their different scales, these modes are nearly identical, differing just by the high frequency component of $G$.

\section{Parameter inference using TTV modes} \label{sec:Mode_recon}

We distinguish between the known and unknown parameters that give rise to TTVs. The period of each planet, and the timing of its first transit are known by a linear fit to the list of transit times, i.e., the linear ephemeris. The planets' masses and eccentricities are unknown and are to be inferred from the TTV signal. We describe our parameter inference procedure using the TTV modes.

\begin{enumerate}
\item Perform a linear fit of the transit times to estimate the periods $P$ and $P'$. Given the (inner planet's) transit times, $t^{tr}_i$, find $P$ and $\tau_0$ that minimize the $\chi^2$ error between the linear ephemeris, $i P + \tau_0$ and the measurements, $t^{tr}_i$. Repeat for the outer planet, finding $P'$ and $\tau_0'$.

\item Given the constant terms in the linear fits, estimate $\Delta \theta_0$, the angle between the planets, assuming circular orbits, at the time of the first transit.

\item Generate the TTV modes given $P'/P$ and $\Delta \theta_0$. This can either be done using analytical TTV formulae \citep{AD_first_2016}, or by performing three N-body orbital integrations, as described in section \ref{sec:contruct_modes}. \label{item:Generate_modes}
\item We define the \textit{modes matrix}
\begin{eqnarray}
\mathbf{M} = 
\begin{bmatrix}
G_1 & F^x_1 & F^y_1 \\ \vdots & \vdots & \vdots \\ G_N & F^x_N & F^y_N
\end{bmatrix} \,,
\end{eqnarray}

where $N$ is the number of available transits in the data. If $\mathbf{y}$ is the vector of TTV measurements,
\begin{equation}
	y_i = t^{tr}_i - ( iP + \tau_0 ) \,,
\end{equation}
the physical parameters that minimize the $\chi^2$ error are given by
\begin{equation}
	\begin{bmatrix}
		\mu' \\ \mu' e_x \\ \mu' e_y
	\end{bmatrix}
= \frac{1}{P} (\mathbf{M}^\intercal \mathbf{M} )^{-1} (\mathbf{M}^\intercal \mathbf{y} ) \,.
\end{equation}
\end{enumerate}

Step \ref{item:Generate_modes} is the most computationally expensive step in our TTV inversion procedure, as it involves 3 numerical orbital integrations. Nonetheless, our method is many orders of magnitude faster than the commonly used numerical methods for TTV inversion \citep{HL_2016}, that require $10^7 - 10^9$ simulations. The orbital integrations we use in constructing the TTV modes can be calculated using \textit{TTVFast} \citep{Deck_TTVFast_2014}, which will expedite the computation times even further.

Given two TTV signals, the masses of both planets may be constrained ($\mu$ from the TTV of the outer planet, and $\mu'$ from inner planet's TTV). The combined free-eccentricity is estimated separately from each planet's TTVs. The eccentricity we finally report is given by the average of the two independent estimates, weighted by their corresponding error estimates.

\begin{figure}[!ht]
\includegraphics[width=1.04\columnwidth]{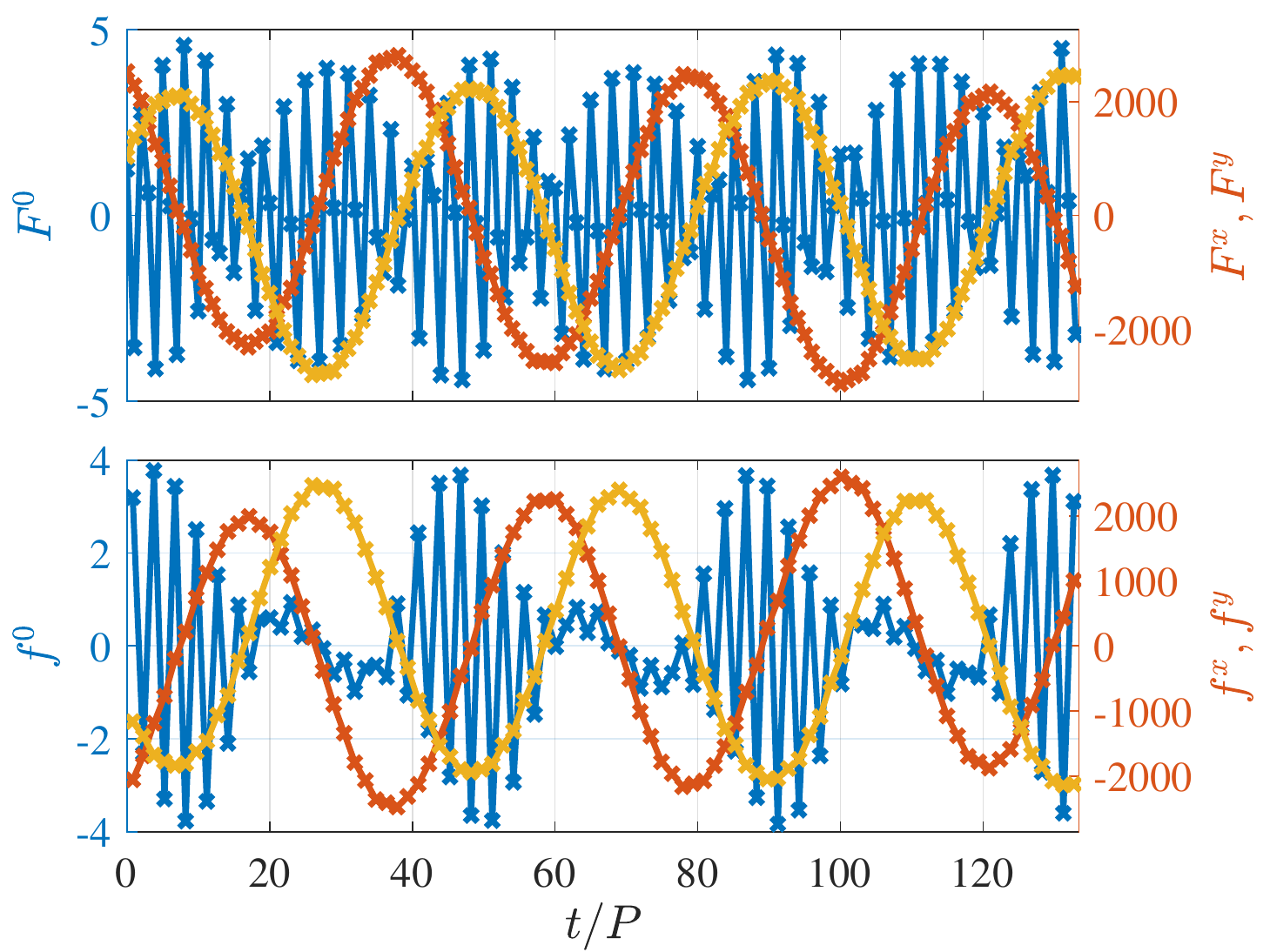}
\caption{TTV modes, calculated for $P'/P = 1.482$, and $\Delta \theta_0 = 9\pi/10$. \textit{Top panel} - the inner planet's TTV modes, \textit{bottom panel} - the outer planet's TTV modes. Left hand vertical axis - the value of $F^0$, right hand - the value of $F^x$ and $F^y$. The \textit{blue curve} is $F^0$, the zero eccentricity term. The \textit{red curve} is $F^x$, the $e_x$ mode. The \textit{yellow curve} is $F^y$, the $e_y$ mode. $F^x$ and $F^y$ are roughly sinusoidal with a timescale $T_{sup}$, and have a phase difference of $\pi/2$. \label{fig:Ortho_Modes_x_0_3}}
\end{figure}

Since the TTV modes are not orthogonal, the inferred physical parameters have correlated errors. As an example, consider a system with an orbital period ratio $P'/P=1.482$, whose corresponding TTV modes are shown in figure \ref{fig:Modes_x_0_3}. When a noisy TTV signal is decomposed as a linear sum of $G$, $F^x$ and $F^y$, there is a strong degeneracy between the coefficients of $G$ and $F^x$, as these modes are nearly parallel. If the higher frequency component of $G$ is not apparent in the data, $\mu'$ and $\mu' e_x$ remain individually unconstrained. This is the so-called mass-eccentricity degeneracy, discussed in \cite{LXW_2012, WL_2013, HL_2014}. When the signal-to-noise ratio is sufficiently high, the mode coefficients are properly constrained. In section \ref{sec:mode_orthogonalization} we discuss the orthogonalization of the TTV modes. This procedure highlights the difference between $G$ and $F^x$, as shown in figure \ref{fig:Ortho_Modes_x_0_3}. The blue curve of figure \ref{fig:Ortho_Modes_x_0_3} gives the high frequency component within the TTV signal that lifts the degeneracy between mass and eccentricity.

\section{Results} \label{sec:Results}

\begin{figure}[ht]
\includegraphics[width=1\columnwidth]{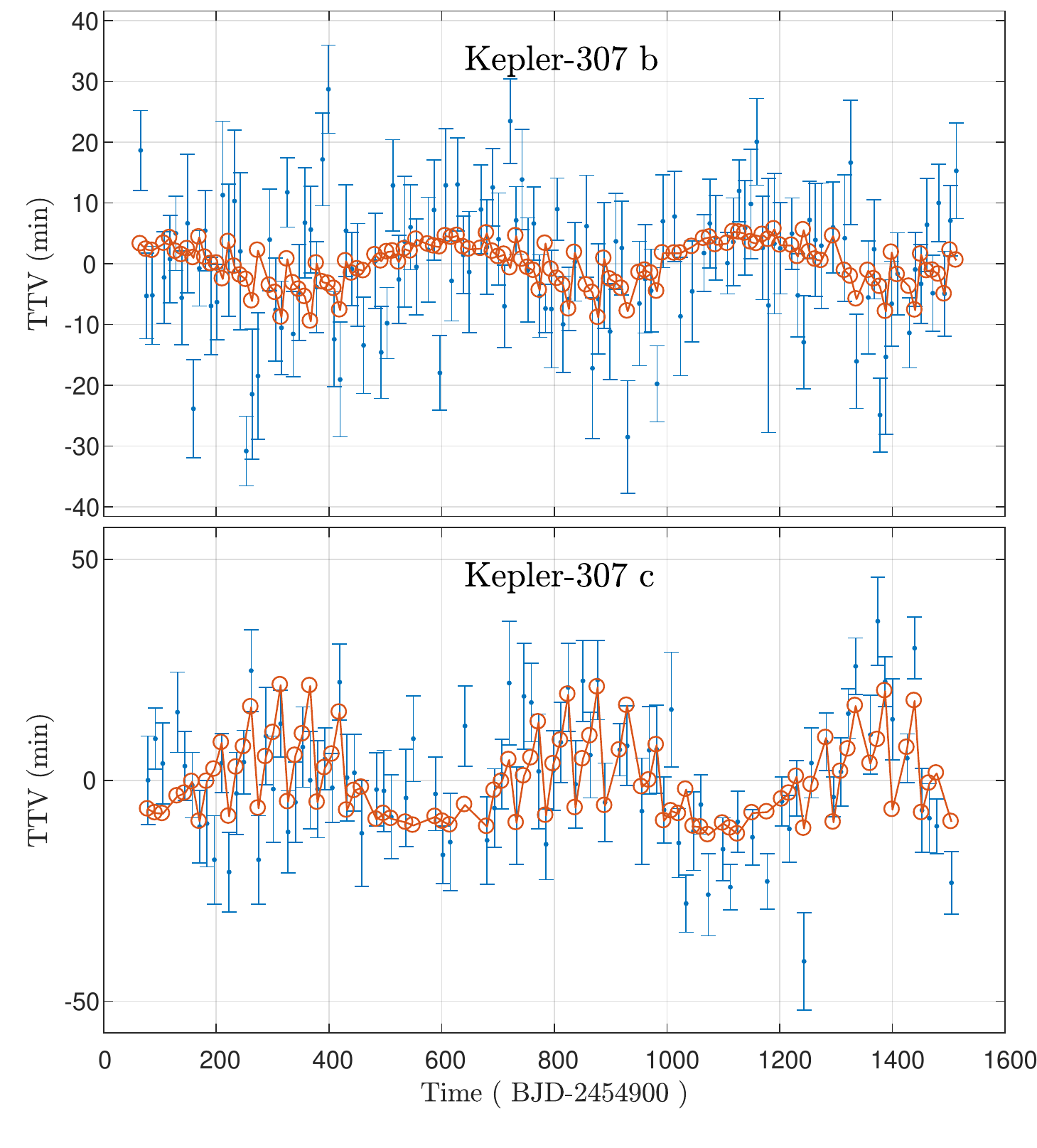}
\caption{TTVs of Kepler-307 b and c. TTVs obtained from \cite{Rowe_Catalog_2015} (for Kepler-307 b) and from \cite{Holczer_Catalog_2016} (for Kepler-307 c) are shown in \textit{blue}. \textit{Red circles} - reconstruction of the signal using the TTV modes, connected by a line. \label{fig:Kepler-307}}
\end{figure}

\begin{figure}[ht]
\includegraphics[width=1\columnwidth]{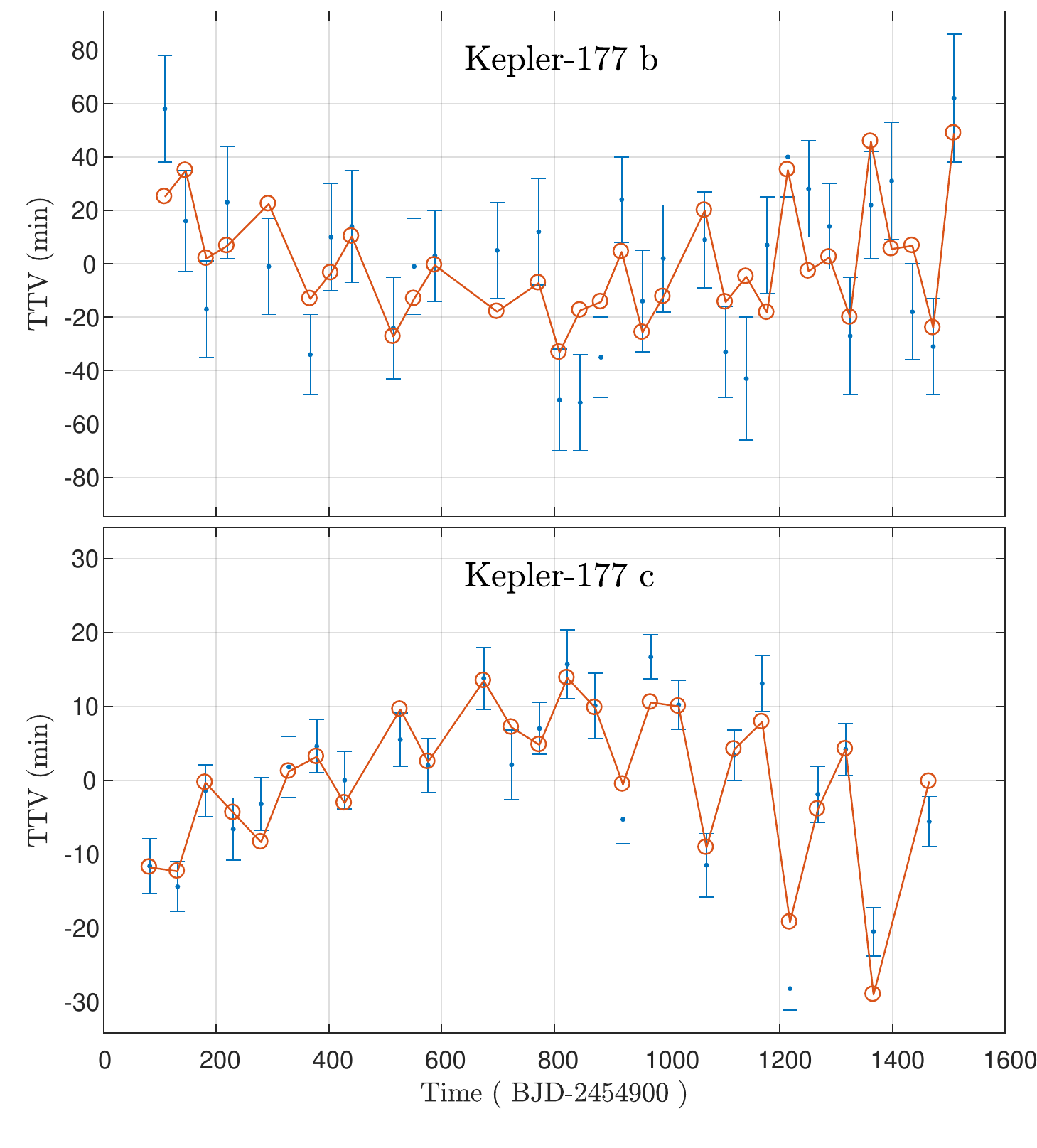}
\caption{TTVs of Kepler-177 b and c (KOI-523.02 and KOI-523.01). TTVs obtained from \cite{Holczer_Catalog_2016} are shown in \textit{blue}. \textit{Red circles} - reconstruction of the signal using the TTV modes, connected by a line. \label{fig:Kepler-177}
}
\end{figure}

\begin{figure}[ht]
\includegraphics[width=1\columnwidth]{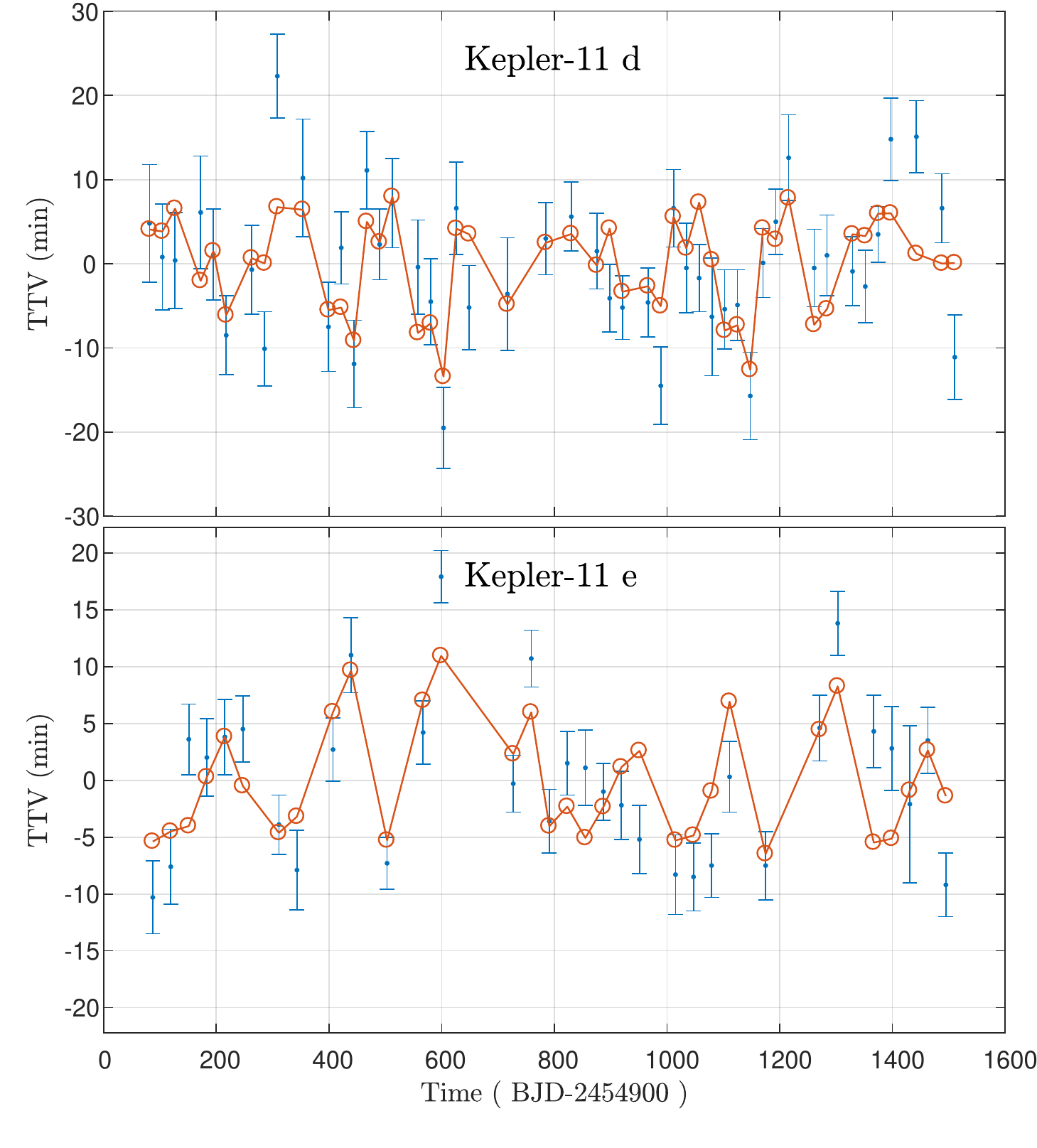}
\caption{TTVs of Kepler-11 d and e (KOI-157.02 and KOI-157.03). TTVs obtained from \cite{Holczer_Catalog_2016} are shown in \textit{blue}. \textit{Red circles} - reconstruction of the signal using the TTV modes, connected by a line. \label{fig:Kepler-11}}
\end{figure}

\begin{figure}[ht]
\includegraphics[width=1\columnwidth]{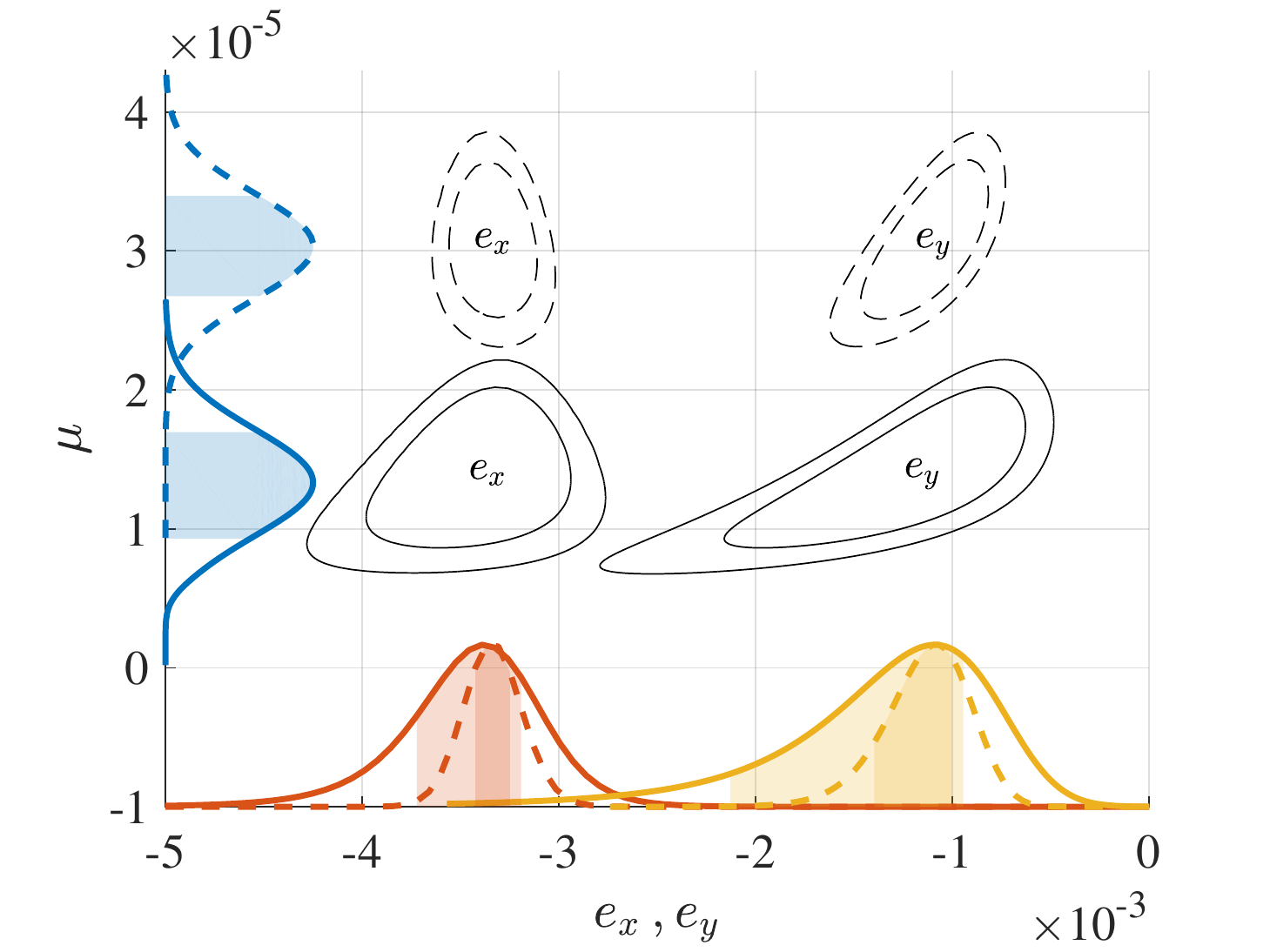}
\caption{The distributions of the physical parameters inferred for Kepler-307 b and c. Filled lines show the parameters inferred from the TTV signal of the inner planet (Kepler-307 b), whereas the dashed lines correspond to the TTV of the outer planet (Kepler-307 c). The contours show the joint-distribution of the perturbing planet's mass and the system's eccentricity component. The contours are labeled by $e_x$ or $e_y$, corresponding to the distribution of the eccentricity component the contours represent. Contours are given at probability values of $0.68$ and $0.87$. Along the axis, the parameters' marginal distributions are shown - \textit{blue} for the planets' relative masses ($\mu$ and $\mu'$), \textit{red} for the $e_x$ component of the combined eccentricity, and $\textit{yellow}$ for the $e_y$ component. The shaded area under the distribution shows the $1\sigma$ range centered at the mean value. Notice that as expected, the eccentricities inferred from the TTVs of both planets are in good agreement, with an overlap of their uncertainty intervals.
\label{fig:Kepler-307_dist}}
\end{figure}

\begin{table*}
\caption{Planet properties for the systems we study. Planet period, radius, stellar mass and their uncertainties (columns 2-4) are taken from the Exoplanet Archive. Columns 5 and 6 were calculated in this work, using the TTV modes. The planet mass uncertainty incorporates the stellar mass uncertainty (column 4), and the error estimate of the planet-to-star mass ratio ($\mu$), given by the least-squares.}
\begin{center}
\begin{tabular}{|c c c c  | c c| c c |}
\hline

{\bf Planet}	& {\bf Period}	& {\bf Radius}	&{\bf Stellar Mass}	& {\bf Mass}	& \bf{ Density} & \bf{Ref. Mass} & \bf{Ref.}\\
	&		 [days]	& $[R_\oplus]$		& $[M_{\odot}]$		& $[M_\oplus]$	& [g/$\text{cm}^3$] & $[M_\oplus]$ & \\ \hline
  
Kepler-307 b & 10.416 & $2.43_{-0.09}^{+0.09}$ & $0.91_{-0.03}^{+0.03}$ & $9.2_{-1.1}^{+1.1}$ & $3.5_{-0.6}^{+0.6}$ & $8.8_{-0.9}^{+0.9}$ & \cite{HL_2017}\\
Kepler-307 c & 13.084 & $2.20_{-0.07}^{+0.07}$ & ---			& $3.96_{-1.33}^{+1.05}$ & $2.0_{-0.7}^{+0.6}$ & $3.9_{-0.7}^{+0.7}$ & \cite{HL_2017}\\ \hline
  
Kepler-177 b & 36.857 & $2.9_{-0.3}^{+1.5}$ & $1.07_{-0.12}^{+0.25}$ & $8.24_{-1.16}^{+2.06}$ & $1.9_{-0.6}^{+2.9}$ & $8.68_{-0.81}^{+0.83}$ & \cite{JH_2016}\\
Kepler-177 c & 49.411 & $7.1_{-0.7}^{+3.7}$ & ---			& $21.1_{-5.5}^{+6.3}$ & $0.33_{-0.13}^{+0.52}$ & $21.91_{-3.92}^{+4.09}$ & \cite{JH_2016}\\ \hline
  
Kepler-11 d & 22.687 & $3.12_{-0.07}^{+0.06}$ & $0.96_{-0.03}^{+0.03}$ & $6.22_{-0.72}^{+0.72}$ & $1.13_{-0.15}^{+0.14}$ & $7.3_{-1.5}^{+0.8}$ & \cite{Lissauer_2013}\\
Kepler-11 e & 31.995 & $4.19_{-0.09}^{+0.07}$ & ---			& $9.4_{-1.3}^{+1.2}$ & $0.70_{-0.10}^{+0.10}$ & $8.0_{-2.1}^{+1.5}$ & \cite{Lissauer_2013}\\ \hline

\end{tabular}
\end{center}
\label{tab:mass_radius}
\end{table*}%

\begin{table*}
\caption{Planet-Pair Eccentricities. The eccentricities we report were found from fitting the data to the TTV modes. Column 2 is the size of the combined free eccentricity of the planet pair. Column 3 ($\varpi$) is the direction of the pericenter with respect to the line of sight. Column 4 gives the value reported by \cite{HL_2017} for the combined eccentricity size.}
\begin{center}
\begin{tabular}{|c c c | c |}
\hline
{\bf Planet pair}	& {$10^3 \cdot e_{free}$}	& {$\varpi (\degr)$} & {Reference $(10^3 \cdot e_{free})$} \\ \hline
  
Kepler-307 b \& c & $3.52\pm 0.16$ & $-161.1\pm 11.1$ & $3.0_{-0.2}^{+0.2}$ \\ \hline
  
Kepler-177 b \& c & $3.39\pm 0.20$ & $173.3\pm 28.5$ &
$2.0_{-0.3}^{+0.2}$ \\ \hline
  
Kepler-11 d \& e & $14.4\pm 5.0$ & $-99\pm 18$ & $9_{-1}^{+1}$ \\ \hline

\end{tabular}
\end{center}
\label{tab:ecc}
\end{table*}%

We demonstrate the use of our inversion method on a few \textit{Kepler} planetary systems. The TTVs obtained from the catalogs of \cite{Holczer_Catalog_2016} and \cite{Rowe_Catalog_2015} for Kepler-307 (KOI-1576), Kepler-177 (KOI-523) and Kepler-11 (KOI-157) are shown in blue in figures \ref{fig:Kepler-307}, \ref{fig:Kepler-177} and \ref{fig:Kepler-11}. The Holczer catalog uses the short-cadence data where available. The TTVs of all six planets (Kepler-307 b \& c, Kepler-177 b \& c, Kepler-11 d \& e) are sufficiently accurate in order to confidently constrain planetary masses and eccentricities with the use of our method.

The red curve in figures \ref{fig:Kepler-307} - \ref{fig:Kepler-11} shows the signal reconstruction using the TTV modes that were calculated for each planet. This reconstruction is the linear combination of the three TTV modes that minimizes the squared differences with the TTV measurements. 

The inferred planet masses and their densities are summarized in table \ref{tab:mass_radius}. The masses we obtain for Kepler-307 b \& c, $M_b = 9.2 \, M_{\oplus}$ and $M_c = 3.96 \, M_{\oplus}$, are consistent with the values reported in \cite{HL_2017}. Note that as they have used different values for the planet's radii, the densities we find are inconsistent with theirs.

Figure \ref{fig:Kepler-307_dist} demonstrates the distributions of the physical parameters inferred for Kepler-307 b \& c. The probability contours of these distributions show the correlation between the physical parameters. We marginalize these distributions and present the distributions of each parameter along the axis of this figure. We overlap the parameter distributions inferred from the TTV of both planets. Since TTVs are sensitive to the combined free-eccentricity, we find similar $e_x$ and $e_y$ values from the TTVs of both planets.

Having used the Holczer TTV catalog for Kepler-177 b \& c, we find planetary masses consistent with those reported by \cite{JH_2016} when they have used the same TTV data. We find the masses of these planets, $M_b = 8.24 \, M_{\oplus}$ and $M_c = 21.1 \, M_{\oplus}$. The bulk density we find for Kepler-177 c is outstandingly low, at about $0.33$ gr/cm$^3$. This density is somewhat higher than the density reported by \cite{HL_2017}, who use the long cadence Rowe catalog \citep{Rowe_Catalog_2015}, and higher than the density reported by \cite{JH_2016}, who use slightly larger planet radius. Note that the fairly large uncertainties in the densities of Kepler-177 b \& c (see table \ref{tab:mass_radius}) are mostly due to the large uncertainties in the planetary radii, taken from the Exoplanet Archive.

Based on our analysis, Kepler-177 c has a surprisingly low bulk density, indicating the presence of a thick atmosphere, which is likely to dominate its volume \citep{Lee_Chiang_2016,Ginzburg_2016}. See further discussion regarding this planet in \cite{JH_2016}

The masses of Kepler-11's six known planets were previously analyzed by \cite{Lissauer_2013} using dynamical models. The masses and densities we report for Kepler-11 d \& e, $M_d = 6.22 \, M_{\oplus}$ and $M_e = 9.4 \, M_{\oplus}$, are in agreement with the uncertainty intervals of \cite{Lissauer_2013}.

Table \ref{tab:ecc} summarizes the inferred eccentricity we find for each of the systems. The combined free eccentricity is given by its magnitude, $e_{free}$ and its direction, $\varpi$. The angle $\varpi$ is the argument of periapsis (of the the combined free eccentricity vector), measured with respect to the line of sight. 

Note that both Kepler-307 and Kepler-177 are quite close to first-order MMR, resulting in a visually distinct sinusoidal feature in the TTV signal, varying on the superperiod in addition to the chopping signal (figures \ref{fig:Kepler-307} and \ref{fig:Kepler-177}). See section \ref{sec:Interpreting_the_ttv_modes} for more on the physical interpretation of the modes.

Planets Kepler-11 d \& e are orbiting close to the $7:5$ second-order MMR. Near second order resonance, second-order terms in eccentricity dominate over the first-order eccentricity terms. Since our linear model does not include any $e_{free}^2$ terms, the eccentricity we infer for this system may be deemed inaccurate, since we assume linear dependence in eccentricity.

Kepler-11 is a system of 6 transiting planets. Nevertheless, we ignore the presence of the additional 4 planets in the analysis of Kepler 11 d \& e. In principle, the TTV of these two planets may contain substantial contributions from the interaction with the additional planets, which we omit. In practice, if the planetary masses are similar and the planet pairs are away from MMR, the TTV is mostly determined by the neighboring planet with the closest orbital period.

\section{Interpretation of the TTV modes} 
\label{sec:Interpreting_the_ttv_modes}

\subsection{Mode orthogonalization} \label{sec:mode_orthogonalization}

As discussed, the TTV modes as defined in equation \ref{eq:gen_ttv_linear} are generally not orthogonal. We note that in systems that are near a first-order MMR, the $G$ and $F^x$ modes are nearly parallel, as is well demonstrated in figure \ref{fig:Modes_x_0_3}. One could orthogonalize the TTV modes using Gram-Schmidt process, to obtain an alternative set of modes. 

In order to elucidate the structure of the TTV modes, we orthogonalize only $G$ and $F^x$, by subtracting from $G$ its projection onto $F^x$, and define
\begin{equation}
	F^0_i = G_i - A F^x_i \,,
\end{equation}
where
\begin{equation}
	A = \frac{\sum_i F_i^x G_i}{\sum_i (F_i^x)^2} \,.
\end{equation}

The general TTV expression (equation \ref{eq:gen_ttv_linear}) then reads
\begin{equation} \label{eq:Ortho_modes}
	ttv_i = \mu' P \left[ F_i^0 + (e_x + A) F_i^x + e_y F_i^y\right] \,.
\end{equation}

In figure \ref{fig:Ortho_Modes_x_0_3} we demonstrate the partly-orthogonalized modes, $F^0$, $F^x$ and $F^y$, calculated for a system with a period ratio $P'/P = 1.482$, similarly to figure \ref{fig:Modes_x_0_3}. 

\subsection{Analogy to previous works}

{The form TTVs take (equation \ref{eq:Ortho_modes}) is reminiscent of the analytical expression given by \cite{HL_2016, HL_2017}. Yet, while \cite{HL_2016, HL_2017} derive their analytical formula under the assumption that the system is near a mean-motion resonance (MMR), the framework we adopt does not assume anything about proximity to MMR. We wish to highlight the mapping between our TTV expressions and the analytical results of \cite{HL_2017}. Note that this mapping fails away from MMR, as their results apply only near MMR.}

{\cite{HL_2016,HL_2017} express a planet's TTV as a sum of harmonic terms. They distinguish between three terms - 'fundamental', 'chopping', and 'secondary' (see equation 1 of \cite{HL_2017} or equation 2 of \cite{HL_2016}).}

{The $F^x$ and $F^y$ modes are analogous to the 'fundamental' TTV component defined in equations 1 and 2 of \cite{HL_2017}. This is a sinusoidal term, whose amplitude increases as the system approaches a first-order MMR. It has the same period as the line of conjunctions period, called the \textit{superperiod}. The contribution from $F^x$ and $F^y$ generally dominates when a system is near a first order mean-motion resonance. In this regime, $F^x$ and $F^y$ converge to sinusoidal functions whose period is the superperiod, sampled at the orbital frequency of the planet. The stripped-down TTV formula given in \cite{LXW_2012}, includes the fundamental component alone. Near resonance, $F^x$ and $F^y$ have a phase difference of roughly $\pi/2$. Thus, the two amplitudes of these modes correspond to the amplitude and phase of the 'fundamental' component.}

{$F^0$ is similar to the 'chopping' TTV term, defined in equations 1 and 5 of \cite{HL_2017}. The amplitude of this term varies smoothly with the period ratio $P'/P$, and does not diverge at quotients of small integers. It has a distinct synodic frequency. Expanded as a sum of frequency components, the chopping term as defined in \cite{HL_2017} excludes the 'fundamental' frequency. This exclusion is analogous to the mode orthogonalization discussed in section \ref{sec:mode_orthogonalization}.}

Note that as we use a linear TTV model, our expression does not include an analog to the 'secondary' component, which is a second-order term in eccentricity. The contribution of this term is important only at systems that are near a second-order mean-motion resonance.

\section{Discussion} \label{sec:Discussion}
TTVs have been measured in a few hundreds of \textit{Kepler} planets to date \citep{Rowe_Catalog_2015,Holczer_Catalog_2016}, and have been used to constrain the masses and eccentricities of a few dozens of planets \citep{HL_2017}. Typically, this is done by using MCMC simulations, searching for the physical parameters most consistent with the observations.

We developed a new method for inverting TTV signals. This method is suited for systems with two transiting planets, with TTV amplitudes that are small compared to the orbital periods. In this regime, TTVs can be expressed as a linear sum of three known functions, accurate to first order in the combined eccentricity and planetary mass. These functions, or \textit{TTV modes}, depend only on the transiting planets' linear ephemerides, which are assumed to be known. Therefore, given a TTV signal, the mass and eccentricity are solved linearly, by decomposing the TTV signal as a sum of the TTV modes.

The modes can be easily constructed numerically, by running 3 orbital integrations of a 3-body system. The mode construction procedure incorporates the possibility of missing transits in the data. If some of the intermediate transit indices are missing, the modes are constructed using the existing indices alone. In this way we avoid systematic differences between the TTV template and the data, such as differences in the linear ephemerides being used.

We demonstrate the use of our method on 6 \textit{Kepler} planets - Kepler-307 b \& c, Kepler-177 b \& c, Kepler-11 d \& e using the TTV catalogs of \cite{Rowe_Catalog_2015} and \cite{Holczer_Catalog_2016}. The masses and combined eccentricities we report are in agreement with masses previously reported by dynamical studies of these planetary systems (see table \ref{tab:mass_radius}).

Our method solves the planetary masses and eccentricities linearly, as opposed to the computationally expensive MCMC simulations. Our approach is valid only in the linear regime, while the numerical calculations include all the additional higher order effects. However, since the planetary masses are small compared to with the stellar mass, and the typical eccentricities are also small, the linear terms are expected to dominate. A possible exception is when a system is near second-order MMR, where the first order terms in eccentricity vanish.

The parameters inferred from the TTV modes can also be used as an initial guess for numerical dynamical investigations. This may help expedite the convergence of MCMC simulations.

Our method is tailored for systems with two planets with measured TTVs. In this case, the TTV of the inner planet is approximated by the expression in equation \ref{eq:gen_ttv_linear}. In the presence of higher multiplicity, the TTV of a given planet will generally depend on the masses and eccentricities of all other planets. As long as the linearity assumption is intact, the TTV of a given planet will be given by the linear sum of the pairwise TTV expressions, each calculated independently. Our method can be therefore extended to analyze the TTVs in systems with higher multiplicities.

Our approach utilizes the linearity of TTVs in order to easily infer the planets masses and eccentricities. The TTV modes can be constructed without the use of analytical formulas, simply by running 3 numerical simulations with arbitrary masses and eccentricities. \cite{AD_first_2016} have provided an analytical TTV expression, accurate to first order in eccentricity (\textit{TTVFaster}). This formula can be used in the construction of the TTV modes, which can then be used to infer the physical parameters of an observed system.
 
\acknowledgments

%




\appendix

\section{Free eccentricities} \label{sec:Appendix_free_eccentricities}
The eccentricity vector of a perturbed planet can be written as
\begin{equation}
\vec{e}_p = \vec{e}_{free} + \vec{e}_{forced} \,,
\end{equation}
where the \textit{forced} eccentricity, $\vec{e}_{forced}$ is induced by the interaction with the companion planets. The \textit{free} eccentricity, $\vec{e}_{free}$ is constant on timescales shorter than the apsidal precession timescale. The eccentricity vector is pointed in the direction of the pericenter, at an angle $\varpi$, measured with respect to the observer's line of sight. 

The eccentricity that enters the TTV expressions is the \textit{combined free eccentricity}. This eccentricity is a weighted difference between the two free eccentricity vectors of the outer and inner planets, with weights determined by the planets' period ratio, $P_{out}/P_{in}$. Different works have used somewhat different normalizations than the one we use \citep{LXW_2012,HL_2016,HL_2017}. In our paper, the combined free eccentricity is defined as
\begin{equation}
	\vec{e} = \vec{e}_{in} - B \, \vec{e}_{out} \,,
\end{equation}
where $B$ is an order-unity coefficient, which depends on $P_{out}/P_{in}$. The value of $B$ can be found from the $f$ and $g$ coefficients listed in the tables of \cite{LXW_2012, HL_2017}. As noted in \cite{HL_2016,HL_2017}, $B \approx 1$, hence the combined free eccentricity vector is roughly the difference between the planets' eccentricities, $\vec{e} \approx \vec{e}_{in} - \vec{e}_{out}$.

\section{Initial conditions for a given free eccentricity} \label{sec:Appendix_Initial_conditions_free_eccentricity}
Consider two coplanar planets of masses $\mu$, $\mu'$, periods $P$ and $P'$ corresponding to the inner and outer planets, respectively. Assume that their initial angular positions are $\theta_0$, $\theta'_0$, and that their initial eccentricities are $\vec{e}_0$ and $\vec{e}'_0$. Assuming that the free and forced eccentricities are small, and that $|\theta_0 - \theta'_0| \approx \pi$, the free eccentricity is given by
\begin{equation}
	\vec{e}_{free} = \vec{e}_0 + \mu' A( \alpha ) \hat{\beta} \,,
\end{equation}
\begin{equation}
	\vec{e'}_{free} = \vec{e'}_0 + \mu A'( \alpha ) \hat{\beta}  \,,
\end{equation}
where $\alpha = P'/P$, and $\hat{\beta}$ is a unit vector pointed at an angle
\begin{equation}
	\beta = \theta_0 + (\alpha - 1)(\theta_0 - \theta'_0 + \pi) \,.
\end{equation}

$A$ and $A'$ are function of $\alpha$, given by
\begin{equation}
	A(\alpha) = \frac{\alpha}{2(\alpha-1) \sin{(\pi \alpha/(\alpha-1))}} \int_{-\pi}^{\pi} d\theta \, \frac{2 \alpha^{2/3} \sin{(\theta \alpha / (\alpha-1))} \sin{(\theta)} - (1-\alpha^{2/3} \cos{(\theta)} ) \cos{(\theta \alpha / (\alpha-1))} }{(1+ \alpha^{4/3} - 2 \alpha^{2/3} \cos{\theta} )^{3/2}} \,,
\end{equation}

\begin{equation}
		A'(\alpha) = \frac{1}{2(\alpha-1) \sin{(\pi/(\alpha-1))}} \int_{-\pi}^{\pi} d\theta \frac{-2 \alpha^{-2/3} \sin{(\theta / (\alpha-1))}\sin{(\theta)} - (1-\alpha^{-2/3} \cos{(\theta)} ) \cos{(\theta / (\alpha-1))} }{(1 + \alpha^{-4/3} - 2 \alpha^{-2/3} \cos{\theta} )^{3/2}} \,.
\end{equation}

\end{document}